\documentstyle[preprint,aps,epsfig,amstex,prl]{revtex}
\begin{document}

\draft

%\twocolumn[\hsize\textwidth\columnwidth\hsize\csname @twocolumnfalse\endcsname
\title{On the constitution of sodium at higher densities}
\author{J. B. Neaton$^*$ and N. W. Ashcroft}
\address{Laboratory of Atomic and Solid State Physics
and Cornell Center for Materials Research,
Cornell University, Ithaca, NY 14853-2501}
\date{October 25, 2000}
\maketitle

\begin{abstract}
Using density functional theory
the atomic and electronic structure of sodium are predicted to depart
substantially from those expected of simple metals for
$r_s <$ 2.48 ($p > 130$ GPa). Newly-predicted phases include those with low
structural symmetry, semi-metallic
electronic properties (including zero-gap semiconducting limiting
behavior), unconventional valence charge density distributions, and even those
that raise the possibility of superconductivity, all at currently achievable
pressures.
Important differences emerge between sodium and lithium at high densities,
and these are attributable to corresponding differences in their
respective cores.
\end{abstract}

\vspace{1cm}


\narrowtext

%]

In their classic papers {\it On the Constitution of Metallic Sodium},
Wigner and Seitz [1] successfully described ground state properties of sodium
invoking the notion of a nearly-free electron metal, one
whose valence electrons are only weakly perturbed by an underlying
periodic arrangement of ions.  An assumption central to their conclusions
(and important to the developing concept of the {\it simple metal})
was that the tightly bound core electrons remain largely unaltered with respect
to those in the free atom. At its stabilizing (atmospheric) density, this
assumption is valid  because the ion cores occupy an exceedingly small fraction
of the overall volume of the solid. But as we show in this Letter, if the
overall density (and hence relative core volume) is increased by pressures that
are currently achievable, the core states come in to play and the electronic
structure of sodium departs quite significantly from this simple metal
paradigm.

Recently it was predicted [1] that the ground state of steadily
compressed lithium, also a prototypical ``simple'' metal, would be
unstable to less symmetric, {\it semi-metallic} structures beginning at
pressures near 50
GPa; and remarkably, it was suggested that the tendency toward less metallic
behavior with increasing pressure would culminate in a nearly-insulating,
{\it paired} ground state
[2] at around 100 GPa.  Two groups, using diamond-anvil cells,
have since reported that lithium gradually loses its Drude-like luster
under compression, first becoming grey and then finally black
(i.e., strongly absorbing) at around 50 GPa [3,4].
Another group, using dynamical shock methods, observed
a declining conductivity with increasing density [4],
broadly consistent with these results.
Recent structural measurements by Hanfland {\it et al.} [6]
find a complex cubic, 16-atom unit cell with space group $I\bar{4}3d$
in lithium near 50 GPa and 200 K. Interestingly enough 
although Ref. [6] reports a
decline in reflectivity against diamond it is suggested that the general
characteristics remain notably metallic.  The companion
first-principles calculations [6] also predict near-insulating
behavior with minimal optical reflectivity at higher pressures
and a subsequent transition to the paired {\it Cmca} phase [2] above 150 GPa.
Though the possibility of broken electronic symmetry states in the light
alkalis has been raised earlier [7], such states are of a very different
physical
character from what is now being found, especially as we show for sodium.

The striking departure from the simple, nearly-free electron-like
behavior conventionally associated with lithium under standard conditions
is attributed in Ref. [2]
to an electron-ion interaction (or pseudopotential) which, on
the length scales of relevance, becomes
increasingly repulsive with density; it originates
with the combined effects of Pauli exclusion and orthogonality.
Given the novel structural and electronic properties predicted to result from
these growing effects under compression, it is natural to examine
the stability of other metals at higher densities often regarded as ``simple''.
In this Letter we report results of first-principles calculations indicating
that some low-coordinated structures first predicted for lithium [2,6] are also
found to be favored in sodium beginning near 130 GPa.  Further, as in
lithium, these
unexpected transitions are accompanied by significant departures from
nearly-free electron-like behavior and should be accessible through optical
response. Differences in the cores of lithium and sodium result in
complementary
differences in ground state atomic and electronic structures; yet sodium is
also
predicted to finally take up the paired {\it Cmca} phase with nearly zero-gap
semiconducting electronic properties.

To study the structural and electronic behavior of sodium over a wide range
of densities, we use a
plane-wave implementation [8] of density functional theory
within the local density approximation [9]
and the projector-augmented wave (PAW) method [10, 11].
As the equation of state will be affected by increasing core overlap at
high densities, we treat the 3$s$, 2$p$, and 2$s$ electrons
as valence and incorporate only the tightly-bound 1$s$ state into an
effective frozen core.
Previous first-principles calculations of standard low- and high-pressure
structural and electronic properties of dense
sodium [13-16]
appear to be limited to the monatomic lattices BCC, FCC, and HCP.
Expermentally it is well known that sodium undergoes a martensitic
transformation
below 35 K, possibly to the 9R phase (as in lithium) [17],
illuminating the role of dynamics in its stabilization at
one atmosphere; our static calculations predict that HCP is very
slightly favored over BCC, FCC, and also 9R at one
atmosphere, but by energies less than 1 meV$/$ion (quite small
compared with dynamical energies and about equal to the accuracy of our
calculations) and consistent with previous work [15].
The equilibrium density we calculate for Na in both the FCC and BCC
structures ($r_s = 3.78$) is also in excellent agreement with previous LDA
results [14, 15], overestimating that found in experiment
($r_s = 3.93$ [18]) by slightly more than 10\%.
(Here $r_s = {(3 V / 4\pi
N_e})^{1 / 3}$, where $V$ is the volume (a.u.$^3$) per ion and $N_e$
corresponds to one
electron/atom throughout.)
Above 5 GPa the BCC structure emerges with lowest enthalpy, and then
remains stable
after an additional 3-fold reduction in volume. It has generally been
expected that BCC should eventually be preferred at the highest densities (the
``free electron''
limit) because of its favorable Madelung energy, and while it might be
suggested that this limit could be reached early for sodium with minimal
compression,
we find instead that BCC is actually unstable to FCC for $r_s \leq 2.66$
($p \geq 71$ GPa), again consistent with the results of a previous study
[16].

Under atmospheric conditions,
the ground state electronic structure of sodium therefore
conforms to the classic nearly-free electron system of Wigner and Seitz [1]
having a single valence electron and an equilibrating density where the core
volume is certainly an exceedingly small
fraction of the cell volume. The pseudopotential for sodium at this density
is quite well approximated as local,
and under moderate compression a purely local picture also continues to
suffice:
the primary Fourier component of the pseudopotential
$V_G$ diminishes relative to the growing bandwidth.
But beyond an approximately 3-fold compression in volume
this simple interpretation no longer holds.
In Fig. 1 we plot the band structure of FCC sodium at $r_s = 2.3$ and
compare it with that of lithium at the same density.
The bands near points $K$, $W$, and $L$ of Na are quite different from
those of Li, and, interestingly, the Fermi surface does remain nearly
spherical in Na at
these elevated densities. Further, the site-projected
partial density of states (DOS) of Na has a growing
$d$ character; in Li, only $s$ character and $p$ character are
observed at $r_s = 2.3$ [19, 20].
However despite these differences (which, as we will emphasize,  are related to
the different electronic configurations of their cores), the energy bands of
sodium and lithium are both appreciably less free electron-like at high
densities:
both exhibit sizable gaps at $X$ and, in contrast to the predictions of the
nearly-free electron model,  these gaps are now {\it growing} in magnitude with
increasing density.

For lithium the departure from nearly-free electron
behavior has evidently been predicted to
result in relatively low-coordinated and open atomic structures [2,6].
Likewise for sodium when the same non-Bravais lattices are in fact considered,
we find that it is also unstable to the formation of these {\it
lower-coordinated} phases for densities greater than $r_s = 2.48$.  A plot of
their enthalpies versus pressure appears in Fig. 2.
The newly-observed cubic $I\bar{4}3d$ phase in lithium [6]
is also found to be favored over FCC in Na at $p \sim 130$ GPa ($\sim$
3.5-fold compression):
in it, each sodium atom has essentially {\it five} neighbors (three at a
distance of 2.18 {\AA} and two at
2.23 {\AA} at a density of $r_s=2.4$) [21]. Before reaching
$r_s = 2.3$, however, our
calculations predict that sodium should transform to Cs-IV at $p
\sim 190$ GPa ($r_s = 2.32$).
The Cs-IV phase, in which each ion is coordinated by four others
[20, 23], can be obtained by a tetragonal distortion of the diamond
structure along [001].
Finally, for densities above $r_s = 2.24$ ($p \sim 280$ GPa), a transition to
{\it Cmca} is predicted, and we find this phase to be favored to the highest
pressures allowed by our PAW potentials ($p \sim 900$ GPa). In each primitive
cell of {\it Cmca},
two {\it pairs} of ions lie in adjacent (100) planes displaced with respect
to one another
by half of a lattice vector in the [010] direction [2, 24].
These predictions are clearly limited by our choice of structures,
and for this reason experiment will again play a crucial role (as in lithium
[6]). It is clear from Fig. 2, however, that simple, monatomic structures are
not stable above 130 GPa [16].
Going beyond LDA we find that gradient corrections [26] result in an expanded
equilibrium volume and shift the Cs-IV to {\it Cmca} transition to
$\sim$260 GPa,
but do not change transition pressures at lower densities.
In addition, spin polarization effects
are found to be insignificant.

Sodium remains metallic in both the $I\bar{4}3d$ and Cs-IV phases,
and notably the transition into each is accompanied by additional
shifts of $p$ and $d$ states below the Fermi energy.  The
Cs-IV phase was not predicted
to precede the {\it Cmca} phase in lithium, and the difference here
may be understood by observing that the $d$ states descend under compression
in sodium (as in dense Cs [25]) but not in lithium
(where they remain far above the Fermi energy to the highest
pressures theoretically examined [20].)
Once in {\it Cmca},
sodium becomes increasingly less metallic as the density rises; the total
and site-projected
DOS are plotted in Fig. 3 for the predicted {\it Cmca} phase near the
highest density
achieved, and most strikingly, at these extremely high pressures (above
$800$ GPa), sodium,
as with lithium before it, approaches a zero-gap semiconducting
phase despite considerable differences in core physics. As a direct result
of the
minimal DOS at the Fermi energy, dense sodium should become increasingly
absorbing in the visible
and its resistivity should increase dramatically [27].
Once again the combined effects of Pauli exclusion and
orthogonality result in an increasing valence electron
density in the {\it interstitial} regions with
density;  we observe that the valence charge density
in $I\bar{4}3d$, Cs-IV, and {\it Cmca} is {\it minimal near and between the
ions} (in regions of maximum core overlap) and {\it maximal in the interstitial
regions} [23]: the valence electrons are evidently forced away from the
cores (and regions of significant core overlap, i.e., the regions
between neighboring ions) into the roomier
interstitial space, resulting in significant benefits for both kinetic and
exchange energies.

The exclusionary effects mentioned above, arising from core overlap, and also the
nature of the cores themselves, strongly influence the atomic and
electronic structure
of compressed sodium, which we find to have growing $p$ and $d$
character with density. At low densities large kinetic energy
costs prevent these bands from dropping
below the Fermi energy. But at higher densities rising core overlap
will favor less symmetric charge distributions, and the associated Bloch
wave functions
have an angular momentum character that depends on the details of core
itself.
Sodium's core, for example, contains both $s$ and $p$ states.
Bloch states with these angular momentum components
(i.e., 3$s$ and 3$p$) are largely excluded from the core region
(where their attraction to the bare nuclear potential $-11e/r$ is greatest),
raising their electrostatic energy.  Bloch states with $d$ character
in sodium (and $p$ character in lithium, as its core contains only the 1$s$
state), however, {\it are} able to sample the full nuclear potential.
(The more asymmetric 3$p$ states in sodium will have a
lower kinetic energy and therefore continue to be important as well.) Thus
in sodium
the 3$d$ bands rapidly drop and increasingly hybridize with the 3$p$ bands
under sufficient compression, even though they remain
nominally above the unoccupied 3$p$ bands at normal densities.
Similarly the 2$p$ bands descend in lithium [19, 20]
because they are absent from its core.

We have seen here that at the relevant length scales the electron-ion
interaction in sodium
is no longer weak under compression
(again as in lithium), and its putative ``simple'' metallic behavior at
low densities appears to be an accident of the
relative core and unit cell volumes present at one atmosphere.
Whether exclusionary effects
actually culminate in an eventual transition to
an insulating state [7] in sodium remains
to be resolved by future experiments.
Since the interstitial charge build-up is a consequence of general quantum
mechanical arguments, the underlying effects revealed here should be evident in
other elements and compounds under sufficiently large compressions.
In addition, the dramatic changes in the atomic and
electronic structure revealed here and in Ref. [2] indicate a large
electron-phonon interaction, enhancing the possibility for observation of
a significant superconducting transition temperature (a possibility also
noted for lithium in Ref. [2]). Up to this point,
superconductivity has not been observed in
sodium at normal densities; but a comprehensive study of its
pressure dependence--both experimentally and theoretically--remains
to be performed. Finally since the calculations presented
above are carried out for static lattices, the temperature dependence of the
structural transformations predicted here will also be a matter of considerable
experimental interest.

We gratefully acknowledge K. Syassen for communicating results prior to
publication, and we thank M. P. Teter and A. Bergara for useful
discussions.  We are indebted to G. Kresse for generating the
PAW potentials and, with J. Hafner, for also providing the VASP code.
This work was supported by the National Science
Foundation (DMR-9988576). This work made use of the Cornell Center for
Materials Research Shared
Experimental Facilities, supported through the National Science Foundation
Materials Research Science and Engineering Centers program (DMR-9632275).

\begin{figure}
\centering
\epsfig{figure=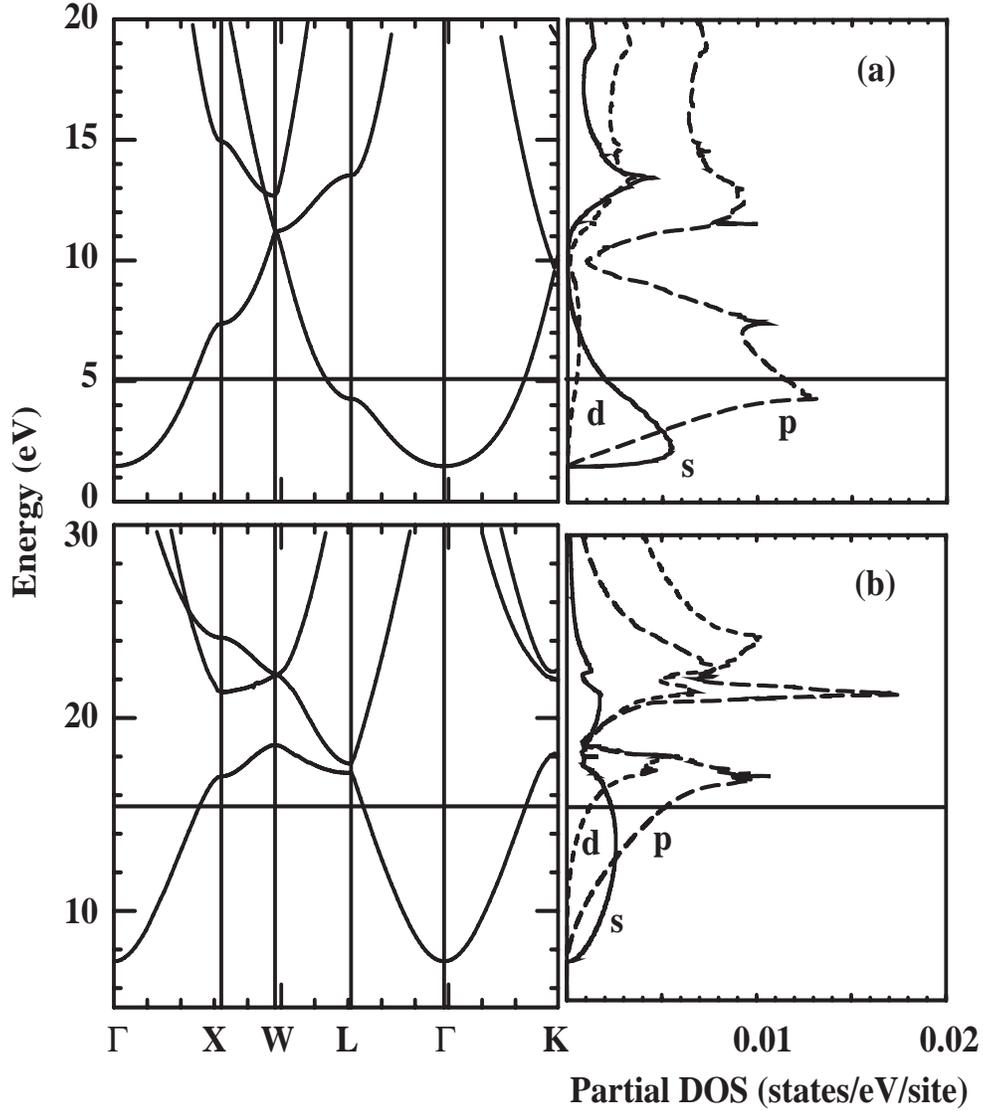,height=15cm,width=13cm}
\caption{Band structures and partial densities of states ($s$, $p$, and $d$)
at $r_s = 2.3$ for FCC (a) lithium ($p \sim 60$ GPa) and (b) sodium ($p
\sim 245$ GPa). Solid horizontal lines denote the Fermi energies. The free
electron value for the Fermi energy at this density is 9.47 eV. The 2$s$ and
2$p$ bandwidths in sodium at this density are 1.25 and 3.5 eV, respectively;
the 1$s$ bands in lithium are 0.7 eV wide.}
\end{figure}

\begin{figure}
\centering
\epsfig{figure=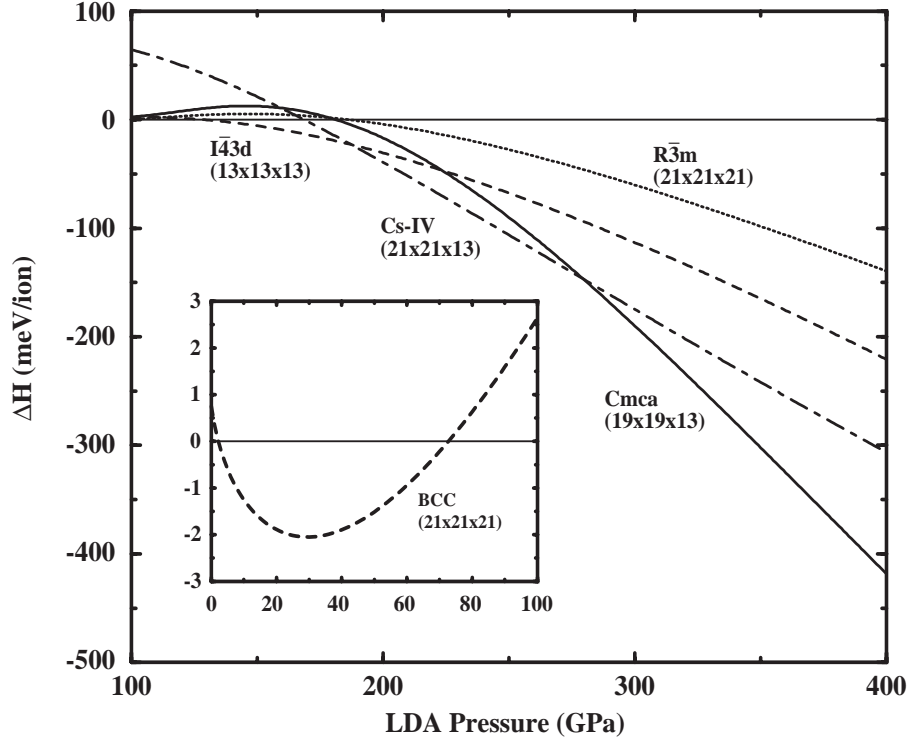,height=10cm,width=12cm}
\caption{Enthalpy difference $H - H_{FCC}$ (meV) vs. pressure $p$
(GPa) for competitive structures of dense Na. The inset illustrates
the calculated low pressure BCC-FCC transition described in the text.
The A7, $Fmmm$, simple cubic, $\beta$-Sn, and
9R phases (all lower in enthalpy than FCC above 250 GPa but
uncompetitive with {\it Cmca}) are omitted from
the figure for clarity.  $R\bar{3}m$ is the space group of the primitive
rhombohedral
phase (see [6]). The Monkhorst-Pack
${\bf k}$-meshes appear in the legend and resulted in convergence of the
total energy to $\sim 1$ meV/ion.  For densities below $r_s = 2.4$ we use a
PAW potential with a 37 Ry plane wave cut-off and maximum core radius of
1.16 {\AA};
for higher densities the distance between neighbors becomes exceedingly small
and we use a potential having a 90 Ry cut-off and maximum core radius
of 0.76 {\AA}. (Results at low densities are independent of the potential.)
$E(V)$ is well fit to $\sum_{n=-2}^2$ $a_{n} V^{n/3}$.
Previous static compressions of Na reached only 30 GPa [19].}
\end{figure}

\begin{figure}
\centering
\epsfig{figure=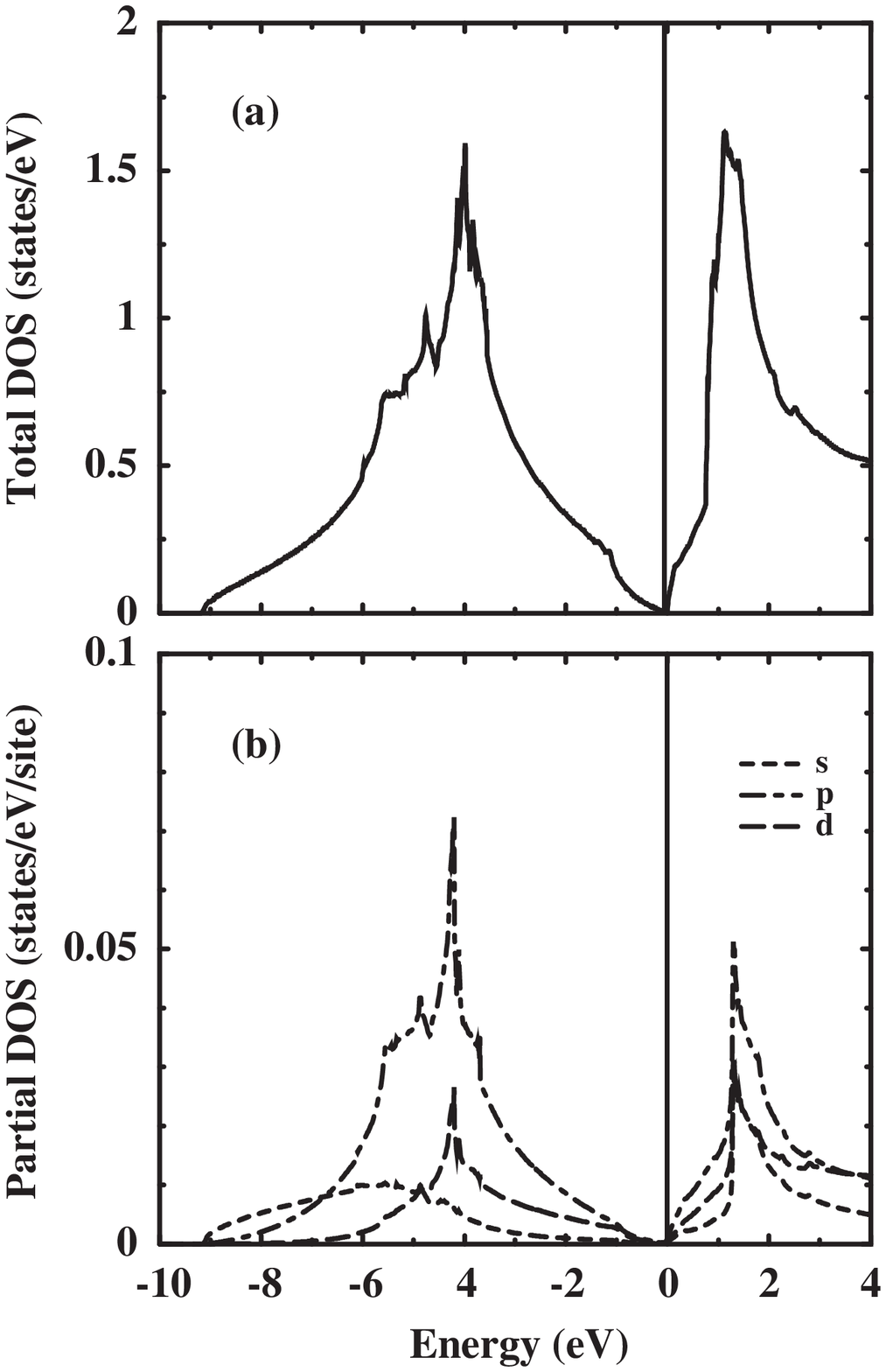,height=16cm,width=10cm}
\caption{Valence band (a) total and (b) partial
density of states of Na in the {\it Cmca}
structure at $r_s = 1.9$ ($\sim$880 GPa). The
partial wave $l$ character of the DOS is primarily $p$, but
there is also a significant amount of $d$ character
which is growing with increasing density.
At lower pressures ($p \sim$ 300-600), the DOS also exhibits a minimum
at the Fermi energy (though it is not zero, as it nearly is above);
the state density at the Fermi energy is found to steadily decline
with increasing pressure.
A solid black vertical line denotes the Fermi energy.
The $l$-projected partial DOS are determined within a 0.75 {\AA} sphere.}
\end{figure}


\begin{references}
\bibitem{ft:liphase}
The $I\bar{4}3d$ phase is
related to BCC by an internal parameter $x$ [6],
which we find to vary from 0.03 at $r_s = 2.5$ to 0.057 at
$r_s = 2.3$ (the BCC value being $x=0$).
\end{references}
\end{document}